\documentclass[aps, reprint, superscriptaddress, nofootinbib]{revtex4-1} %reprint or preprint%
\usepackage[dvipdfmx]{graphicx}
\usepackage{dcolumn}
\usepackage{amsmath}
\usepackage{amssymb}
\usepackage{xcolor}
\usepackage{bm}
\usepackage{braket}
\usepackage{txfonts}

\newcommand{\argmax}{\mathop{\rm argmax}\limits}

\begin{document}
%\preprint{HEP/123-qed}
\title{Phase transitions from linear to nonlinear information processing in neural networks}
\author{Masaya Matsumura}
\author{Taiki Haga}
\email[]{taiki.haga@omu.ac.jp}
\affiliation{Department of Physics and Electronics, Osaka Metropolitan University, Sakai-shi, Osaka 599-8531, Japan}
\date{\today}

\begin{abstract}
We investigate a phase transition from linear to nonlinear information processing in echo state networks, a widely used framework in reservoir computing.
The network consists of randomly connected recurrent nodes perturbed by a noise and the output is obtained through linear regression on the network states.
By varying the standard deviation of the input weights, we systematically control the nonlinearity of the network.
For small input standard deviations, the network operates in an approximately linear regime, resulting in limited information processing capacity.  
However, beyond a critical threshold, the capacity increases rapidly, and this increase becomes sharper as the network size grows.  
Our results indicate the presence of a discontinuous transition in the limit of infinitely many nodes.  
This transition is fundamentally different from the conventional order-to-chaos transition in neural networks, which typically leads to a loss of long-term predictability and a decline in the information processing capacity.  
Furthermore, we establish a scaling law relating the critical nonlinearity to the noise intensity, which implies that the critical nonlinearity vanishes in the absence of noise.
\end{abstract}

\maketitle

\section{Introduction}

% Information processing in dynamical systems
Information processing in dynamical systems refers to how these systems encode and transform information over time through their evolving states. 
A dynamical system, defined by its state variables and evolution rules, can be considered as processing information by responding to external inputs. 
A typical example is the brain's neural networks, where neurons interact dynamically to encode sensory inputs, process them through layers of neural activity, and produce cognitive outputs. 
The capacity of a dynamical system to process information depends on the complexity of the responses it can produce in response to a given time series input. 
Understanding the relationship between the information processing capacity of a system and its dynamical properties is crucial, not only for fundamental research but also for practical applications.

% Review of reservoir computing
Reservoir computing is a computational framework that harnesses the information processing capacity of high-dimensional nonlinear dynamical systems, known as reservoirs, to efficiently process time series data \cite{Jaeger-09, Tanaka-19,Nakajima-21, Cucchi-22}.
Unlike traditional neural networks where all layers are trained, reservoir computing keeps the reservoir structure fixed, training only a simple readout layer. 
The reservoir acts as a nonlinear mapping that projects time series data into a high-dimensional state space, where complex patterns and temporal dependencies are more easily separable. 
This approach simplifies the training process and has been effectively applied to tasks such as time series prediction \cite{Pathak-17, Lu-17, Pathak-18, Vlachas-20, Gauthier-21, Zhang-21, Kobayashi-21, Rohm-21, Srinivasan-22, Storm-22, Tanaka-22, Zhang-23} and speech recognition \cite{Verstraeten-05, Skowronski-07, Rodan-11}, using both virtual reservoirs like recurrent neural networks and physical reservoirs like optical and mechanical systems \cite{Appeltant-11, Nakajima-15, Dion-18, Du-17, Moon-19, Toprasertpong-22, Yada-21, Maksymov-22, Larger-17, Rafayelyan-20, Wang-24}.

% Effects of nonlinearity
Nonlinearity is essential for the information processing capabilities of dynamical systems. 
Linear systems generally have limited processing capacity because their responses to inputs are merely scaled and summed versions of the inputs, which implies that they can only handle linearly separable data. 
It has been known that introducing even weak nonlinearity can significantly enhance the information processing capacity of the system \cite{Inubushi-17}.
On the other hand, strong nonlinearity may push the system into chaotic behavior, which undermines the reproducibility of outputs and reduces the system's ability to make reliable predictions. 
For optimal information processing, nonlinearity must be balanced to avoid the extremes of linearity and chaotic behavior. 
It is commonly believed that the information processing capacity is maximized at the edge of chaos, which is the boundary between ordered and chaotic dynamics \cite{Bertschinger-04, Schweighofer-04, Legenstein-07, Busing-10, Toyoizumi-11, Wainrib-13, Aljadeff-15, Livi-17, Carroll-20}.
Nonetheless, the exact influence of nonlinearity on information processing capacity is not yet fully understood, and the optimal balance between linearity and nonlinearity remains debated.

%%%%%%%%%%%%%%%%%%%%%%%%%%%%%%%%%%%%
\begin{figure}[b]
\centering
\includegraphics[width=8.6cm]{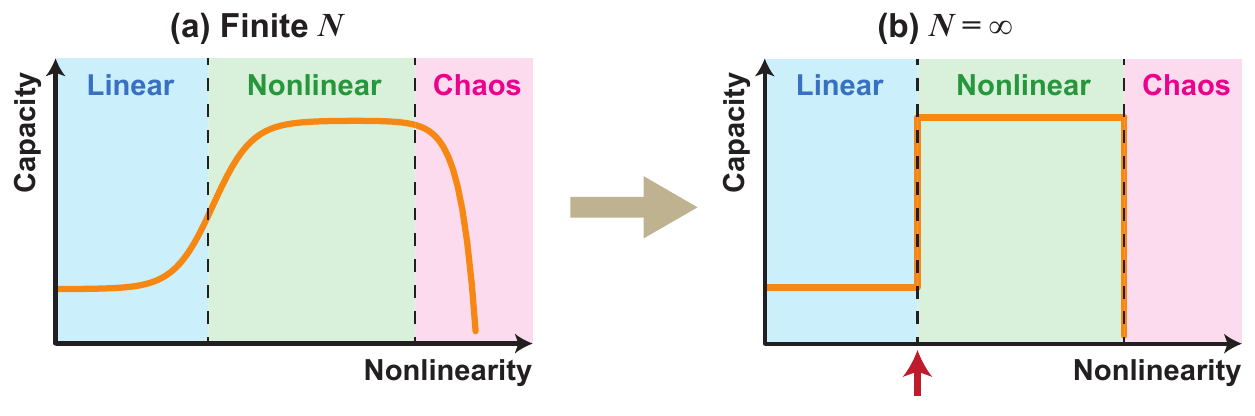}
\caption{Schematic illustration of the information processing capacity as a function of system nonlinearity. 
(a) For a finite number of degrees of freedom, the capacity increases gradually with weak nonlinearity and decreases rapidly when strong nonlinearity induces chaos. 
The question arises as to what happens when the number of degrees of freedom goes to infinity. 
As illustrated in panel (b), we demonstrate a possibility that the capacity undergoes a discontinuous jump at a critical point, indicated by the red arrow.}
\label{fig_transition_schematic}
\end{figure}
%%%%%%%%%%%%%%%%%%%%%%%%%%%%%%%%%%%%

% Transition from linear to nonlinear information processing
In this work, we focus on the transition from linear to nonlinear information processing in the limit of infinite degrees of freedom.
In systems with finite degrees of freedom, this transition from the linear to weakly nonlinear regime is typically smooth and gradual [see Fig.~\ref{fig_transition_schematic}(a)].
However, as the number of degrees of freedom increases, the interactions between individual components can lead to emergent collective behavior. 
In such large systems, even small increases in the nonlinearity parameter may cause the system to abruptly shift from linear behavior to nonlinear collective dynamics. 
This sudden change could lead to a discontinuous transition in the information processing capacity of the system, akin to first-order phase transitions [see Fig.~\ref{fig_transition_schematic}(b)]. 
Understanding these transitions between linear and nonlinear regimes in information processing may reveal novel universality classes in nonequilibrium phase transitions.

% Difference from order-to-chaos transitions
We emphasize that the transitions from linear to nonlinear information processing are fundamentally different from the transitions from order to chaos in complex systems.
The linear to nonlinear information processing transitions are characterized by the ability of the system to perform increasingly complex input-output transformations, with a focus on enhancing computational power. 
In contrast, the order to chaos transitions involve a change in the dynamical behavior of the system, marked by the sign of the maximal Lyapunov exponent.
For effective learning in reservoir computing, it is crucial that the system satisfies the echo state property, which ensures that the response of the system is uniquely determined by the input history \cite{Jaeger-09}.
In the chaotic regime, this property is lost, leading to a significant reduction in information processing capacity (see Fig.~\ref{fig_transition_schematic}). 
While the order-to-chaos transitions in neural networks have been extensively studied over the past decades \cite{Sompolinsky-88, Massar-13, Kadmon-15, Schuecker-18, Kusmierz-20, Engelken-23}, systematic investigations of linear to nonlinear information processing transitions remain largely unexplored.

% What is done in this work
In the following, we demonstrate a phase transition from linear to nonlinear information processing in echo state networks, which are a standard paradigm in reservoir computing using recurrent neural networks.
The reservoir consists of randomly connected nodes perturbed by a small amount of noise. 
The output is computed as a linear combination of the reservoir states, where weights are trained via standard linear regression to minimize the distance between the output and the target data. 
The information processing capacity is measured by the correlation coefficient between the output and the target. 
The nonlinearity of the system is controlled by the standard deviation of the input weights. 
When this value is small, the reservoir states remain small and are minimally influenced by the nonlinearity of the activation function. 
By examining the information processing capacity as a function of the input weight standard deviation, we observe a rapid increase in the capacity from the linear to the nonlinear regime at a critical value, with the transition becoming sharper as the number of nodes increases [see Fig.~\ref{fig_r2_map}(a)-(d)].
This suggests the possibility of singular behavior in the information processing capacity in the limit of an infinite number of nodes.
Additionally, we find that as the noise intensity decreases, the standard deviation of the input weights at which the phase transition occurs approaches zero, implying that even an infinitesimally weak nonlinearity can drive the system into the nonlinear information processing regime.

Here, we clarify our use of the term ``phase transition".
In this study, we use it in a broad sense to describe a collective phenomenon in which an observable exhibits non-analytic behavior as a function of a control parameter \emph{in the limit of infinite system size}.
This definition highlights the distinction between our transition and bifurcation phenomena such as order-to-chaos transitions, which can occur even in finite systems.
In particular, our transition does not necessarily display features commonly associated with conventional thermodynamic phase transitions, such as the divergence of correlation length or correlation time.

This paper is organized as follows.  
In Sec.~\ref{sec:information_processing_capacity}, we introduce a general framework for defining the information processing capacity of dynamical systems.  
By adopting the reservoir computing perspective, the capacity is formulated as a measure of the system's ability to reproduce a given input-output relationship.
In Sec.~\ref{sec:echo_state_network}, we describe the echo state network model perturbed by a small amount of noise.  
We detail the initialization procedure for the recurrent and input weight matrices, and explain the method used to calculate the Lyapunov exponent of the network.
In Sec.~\ref{sec:results}, we present our main results on the information processing capacity.  
The phase diagram shown in Fig.~\ref{fig_r2_map} highlights the existence of distinct linear and nonlinear information processing regimes.
We further analyze how the capacity depends on the noise intensity and network size.
Sec.~\ref{sec:conclusion} concludes the paper with a summary and outlook for future research.
Appendix~\ref{sec:training_procedure} provides the detailed procedure for training the echo state network.
Appendix~\ref{sec:finite_size_scaling_suppl} details a finite-size scaling analysis of the transition.
In Appendix~\ref{sec:ridge_regression}, we examine how the main results are modified when ridge regression is used instead of standard linear regression.
Appendix~\ref{sec:narma_delay} discusses the noise dependence of the information processing capacity for different tasks.
In Appendix~\ref{sec:activation_function}, we examine how the choice of activation function influences the noise dependence of the transition point.
Finally, in Appendix~\ref{sec:round-off_error}, we address the impact of round-off errors in numerical calculations, which become significant when both the noise intensity and input weight magnitude are extremely small.

\section{Information processing capacity}
\label{sec:information_processing_capacity}

In this section, we define the information processing capacity of dynamical systems according to the framework of reservoir computing.
Let $\mathbf{x}(t) = (x_1(t), \ldots, x_N(t)) \in \mathbb{R}^N$ denote the state of a dynamical system at discrete time $t \in \mathbb{Z}$, and let $\mathbf{u}(t) = (u_1(t), \ldots, u_D(t)) \in \mathbb{R}^D$ be the input. 
The time evolution is given by
\begin{equation}
\mathbf{x}(t+1) = f(\mathbf{x}(t), \mathbf{u}(t+1)),
\end{equation}
where $f: \mathbb{R}^N \times \mathbb{R}^D \to \mathbb{R}^N$ is a deterministic map.

The time evolution of the system defines the transformation of the input sequence $\{\mathbf{u}(t)\}$ to the state sequence $\{\mathbf{x}(t)\}$.
The information processing capacity of the system reflects the complexity of this transformation.
To quantify the capacity of the system, we consider the following procedure.
Firstly, a target signal $\mathbf{z}(t) \in \mathbb{R}^M$ is generated via
\begin{equation}
\mathbf{z}(t) = g(\mathbf{u}(t), \mathbf{u}(t-1), \ldots),
\end{equation}
where $g$ is a nonlinear function of the past inputs $\mathbf{u}(s)$ with $s \leq t$.
From the system state $\mathbf{x}(t)$, the output $\mathbf{y}(t) \in \mathbb{R}^M$ is constructed as a linear combination,
\begin{equation}
\mathbf{y}(t) = \mathbf{w}^{\text{out}}_0 + \sum_{i=1}^N \mathbf{w}^{\text{out}}_i x_i(t).
\label{model_output}
\end{equation}
The output weights $\mathbf{w}_i^{\text{out}} \in \mathbb{R}^M \: (i=0,1, \ldots, N)$ are optimized to minimize the total squared error over a time window of length $T$,
\begin{equation}
E[\{\mathbf{w}_i^{\text{out}}\}] := \sum_{t=1}^T \|\mathbf{y}(t)-\mathbf{z}(t)\|^2.
\end{equation}

After optimizing the output weights $\mathbf{w}^{\text{out}}_i$, we compare the output sequence $\mathbf{y}(t)$ to the target $\mathbf{z}(t)$.
We measure the closeness of $\mathbf{y}(t)$ and $\mathbf{z}(t)$ using the square of the Pearson correlation coefficient, 
\begin{equation}
r^2 = \frac{\text{cov}(\mathbf{y}, \mathbf{z})^2}{\text{var}(\mathbf{y}) \text{var}(\mathbf{z})}, 
\label{r2_definition}
\end{equation}
where $\text{cov}$ and $\text{var}$ denote covariance and variance, respectively.
A value of $r^2=1$ indicates a perfect match with the target, while $r^2=0$ means the model fails to reproduce the target.
In this study, we refer to $r^2$ as the information processing capacity associated with the target function $g$.
Note that, in evaluating $r^2$, it is essential to use new input and target sequences that are independent of the training data.
This prevents overfitting and ensures that the computed capacity reflects the generalization performance of the system.

The value of $r^2$ reflects the system's ability to simulate the input-output relation defined by the target function $g$. 
A higher value of $r^2$ indicates that the system effectively transforms the input history $\mathbf{u}(0), \ldots, \mathbf{u}(t)$ into a rich internal state $\mathbf{x}(t)$ that supports accurate reconstruction of $\mathbf{z}(t)$.
Conversely, as the complexity of the target function $g$ increases, the achievable $r^2$ typically decreases.
Note that the definition of information processing capacity adopted here is a simplified version of the one introduced in Ref.~\cite{Dambre-12}, where a set of target functions forming an orthogonal basis in the function space of $g$ is used to define a family of capacities.

\section{Echo state network}
\label{sec:echo_state_network}

We consider a recurrent neural network with $N$ nodes, where the state of the network at time $t \in \mathbb{Z}$ is represented by a column vector $\mathbf{x}(t) = (x_1(t), ..., x_N(t))$.
The time evolution of $\mathbf{x}(t)$ is given by 
\begin{equation}
\mathbf{x}(t+1) = h( W \mathbf{x}(t) + \mathbf{v} u(t+1) + \boldsymbol{\xi}(t+1) ), 
\label{network_evolution}
\end{equation}
where $W \in \mathbb{R}^{N \times N}$ is the recurrent weight matrix, $\mathbf{v} \in \mathbb{R}^N$ is the input weight, and $u(t) \in \mathbb{R}$ is the input signal. 
The function $h(\mathbf{x})$ is an element-wise activation function.
We also include an independent Gaussian noise $\boldsymbol{\xi}(t) \in \mathbb{R}^N$ with mean zero and standard deviation $\sigma_{\xi}$, 
\begin{equation}
\langle \xi_i(t) \xi_j(s) \rangle = \sigma_{\xi}^2 \delta_{ij} \delta_{ts}.
\end{equation}
This noise represents thermal fluctuations or measurement errors in physical reservoir computing.
We refer to networks using the activation function $h(\mathbf{x})=\mathbf{x}$ as linear networks, and those using $h(\mathbf{x})=\tanh(\mathbf{x})$ as nonlinear networks.

The output $y(t) \in \mathbb{R}$ is given by Eq.~\eqref{model_output} with $M=1$.
In an echo state network, the matrices $W$ and $\mathbf{v}$ remain fixed, while only the output weights $w^{\text{out}}_i$ are trained via linear regression.
The information processing capacity $r^2$ is calculated according to the procedure described in Sec.~\ref{sec:information_processing_capacity}.
In the following, we mainly use a random binary input $u(t) \in \{0, 1 \}$ and the XOR function as the target function, 
\begin{equation}
z(t) = u(t) + u(t-k) \quad (\text{mod} \: 2), 
\label{XOR}
\end{equation}
where $k > 0$ sets the delay for the past input's influence.
The details of the training procedure are presented in Appendix \ref{sec:training_procedure}.
It is important to note that a linear network with the identity activation function $h(\mathbf{x}) = \mathbf{x}$ yields $r^2 = 0$, since the XOR function defined in Eq.~\eqref{XOR} is not linearly separable.
That is, the XOR target cannot be represented by any linear combination of the input $u(t)$.  

The input weights $\mathbf{v}=(v_1, ..., v_N)$ are drawn from a uniform distribution on $[-\sqrt{3} \sigma_{\text{in}}, \sqrt{3} \sigma_{\text{in}}]$, where $\sigma_{\text{in}}$ is the standard deviation that controls the strength of the input.
In a nonlinear network with the activation function $h(\mathbf{x})=\tanh(\mathbf{x})$, the input standard deviation $\sigma_{\text{in}}$ governs the degree of the nonlinearity: if $\sigma_{\text{in}} \ll 1$, then $|x_i(t)| \ll 1$, making $\tanh(\mathbf{x}) \simeq \mathbf{x}$ effectively linear.
The elements of the recurrent weight matrix $W_{ij}$ are first drawn from a uniform distribution on $[-1, 1]$ with probability $p$, and set to zero with probability $1-p$.
The matrix $W$ is then rescaled so that its spectral radius equals a prescribed value $\rho$.
Throughout our calculations, we fix the connectivity factor $p=0.1$.

We also calculate the Lyapunov exponent of the network to distinguish between ordered and chaotic dynamical regimes \cite{Storm-22}.  
For convenience, we define the pre-activation state as
\begin{equation}
\mathbf{r}(t+1) = W \mathbf{x}(t) + \mathbf{v} u(t+1) + \boldsymbol{\xi}(t+1).
\end{equation}
Suppose two initially close states \( \mathbf{x}(0) \) and \( \mathbf{x}'(0) \).  
The Lyapunov exponent \( \lambda \) is defined as the exponential divergence rate between these states:
\begin{equation}
\| \mathbf{x}'(t) - \mathbf{x}(t) \| \sim e^{\lambda t} \| \mathbf{x}'(0) - \mathbf{x}(0) \|.
\end{equation}
The evolution of the difference vector \( \delta \mathbf{x}(t) = \mathbf{x}'(t) - \mathbf{x}(t) \) is given by
\begin{equation}
\delta \mathbf{x}(t+1) = \tilde{X}(t+1) W \delta \mathbf{x}(t),
\end{equation}
where \( \tilde{X}(t) \) is a diagonal matrix with elements \( \tilde{X}_{ii}(t) = 1 - \tanh^2(r_i(t)) \).  
The Lyapunov exponent \( \lambda \) is then calculated as
\begin{equation}
\lambda = \lim_{t \to \infty} \frac{1}{t} \log \frac{\| \tilde{X}(t) W \tilde{X}(t-1) W \cdots \tilde{X}(1) W \delta \mathbf{x}(0) \|}{\| \delta \mathbf{x}(0) \|}.
\end{equation}
In numerical simulations, the value of $\lambda$ is averaged over several random initial vectors $\delta \mathbf{x}(0)$.

\section{Results}
\label{sec:results}

%%%%%%%%%%%%%%%%%%%%%%%%%%%%%%%%%%%%
\begin{figure*}
\centering
\includegraphics[width=\textwidth]{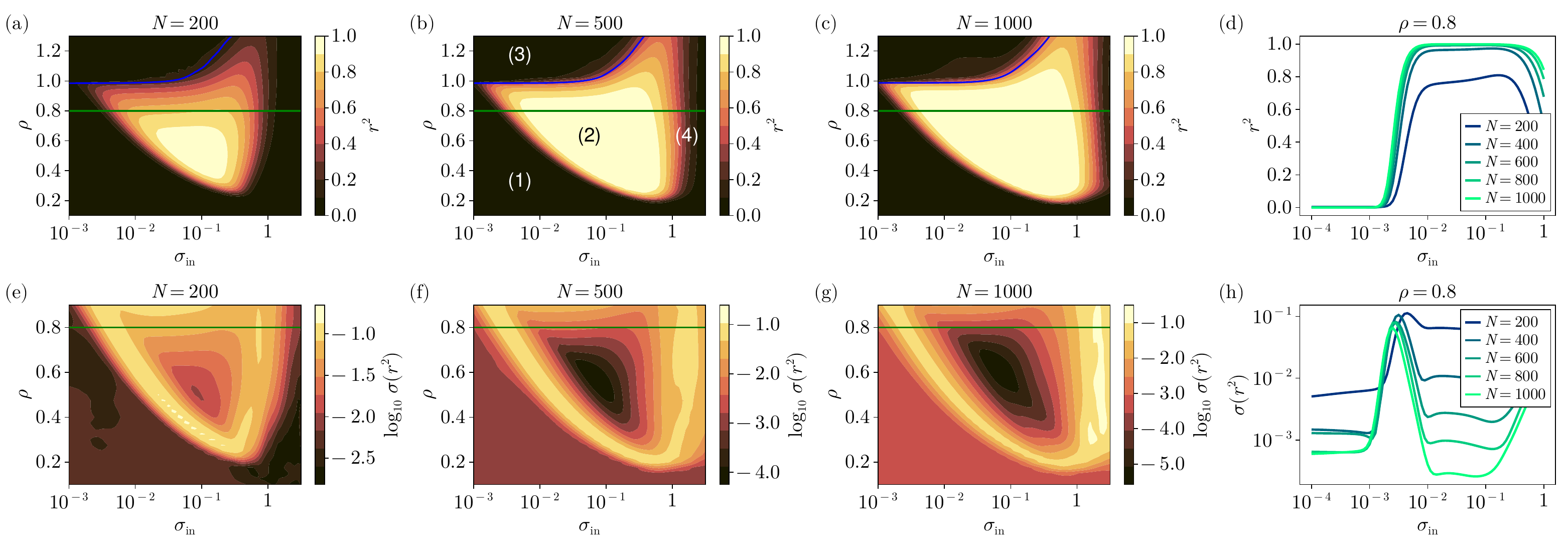}
\caption{Information processing capacity $r^2$ and its fluctuations $\sigma(r^2)$ for an echo state network as a function of the input standard deviation $\sigma_{\text{in}}$ and the spectral radius $\rho$ of the recurrent weight matrix.
Panels (a)-(c) show the averaged capacity $r^2$ for different numbers of nodes, $N=200$, $500$, and $1000$.
The blue solid curves indicate the boundary of the order-to-chaos transition, determined by the Lyapunov exponent.
Panel (d) shows similar data for $\rho=0.8$, which is highlighted by the green solid lines in (a)-(c).
We identify four regimes (1)-(4) in (b). 
Regimes (1) and (2) correspond to linear and nonlinear information processing, respectively, and the boundary between these two regimes becomes sharper as $N$ increases.
Panels (e)-(g) show the logarithm of the standard deviation $\sigma(r^2)$ of the capacity over different network realizations, while panel (h) shows $\sigma(r^2)$ for $\rho=0.8$.
At the transition between regimes (1) and (2), $\sigma(r^2)$ exhibits a pronounced peak, signaling a phase transition.
For all data, the noise intensity is $\sigma_{\xi}=10^{-8}$ and the delay for the XOR task is set to $k=10$.}
\label{fig_r2_map}
\end{figure*}
%%%%%%%%%%%%%%%%%%%%%%%%%%%%%%%%%%%%

\subsection{Phase diagram}

Figures \ref{fig_r2_map}(a)-(c) show the capacity $r^2$ of the nonlinear network as a function of the input standard deviation $\sigma_{\text{in}}$ and the spectral radius $\rho$ of the recurrent weight matrix, for different numbers of nodes $N$.
In the following, the reported values of $r^2$ are averaged over 10 to 50 independent realizations of the network, depending on the network size $N$.
As $\sigma_{\text{in}}$ and $\rho$ increase, the network exhibits stronger nonlinearity.
The blue solid curve near $\rho=1$ indicates the boundary of the order-to-chaos transition, as determined by the condition that the Lyapunov exponent $\lambda$ approaches zero.
Specifically, we plot the contour corresponding to $\lambda = -0.03$, since the Lyapunov exponent in the chaotic regime remains close to zero, and the exact contour for $\lambda = 0$ does not yield a smooth or numerically stable boundary.
Below this blue curve, the Lyapunov exponent is negative, indicating that the network operates in a non-chaotic, ordered regime.  

We identify four phases (1)-(4), as highlighted in Fig.~\ref{fig_r2_map}(b).
In phase (1), characterized by small $\sigma_{\text{in}}$ and $\rho$, the system effectively operates as a linear network, so its capacity is zero for the XOR task.
In phase (2), nonlinearity becomes significant, and the network achieves a high capacity.
In phase (3), where $\rho$ is large, the network enters a chaotic regime with a positive Lyapunov exponent, and the capacity drops to zero.
In phase (4), large $\sigma_{\text{in}}$ drives the state $x_i(t)$ to saturate at $\pm 1$ under the $\tanh$ activation, preventing effective nonlinear transformations.
Our primary focus is the boundary between phases (1) and (2).
Remarkably, as the number of nodes $N$ increases, this phase boundary becomes sharper, as also seen in Fig.~\ref{fig_r2_map}(d).
This observation suggests that the capacity $r^2$ exhibits a discontinuous jump in the limit $N \to \infty$, as illustrated in Fig.~\ref{fig_transition_schematic}.

Figures \ref{fig_r2_map}(e)-(h) show the fluctuation $\sigma(r^2)$ of the capacity across different network realizations.
A pronounced peak appears at the boundary between phases (1) and (2).
This is reminiscent of the divergence of order-parameter fluctuations observed at second-order phase transitions if one considers the capacity $r^2$ as the order parameter.
However, there are notable differences from the conventional phase transition picture.
First, our transition appears to be first-order (discontinuous), rather than second-order (continuous).
Second, the height of the peak of $\sigma(r^2)$ does not diverge in the limit $N \to \infty$.
This bounded behavior arises from the fact that $r^2$ itself is confined to the interval $0 \leq r^2 \leq 1$, in contrast to extensive order parameters in traditional systems, such as magnetization in the Ising model, which scale with system size.
Finally, the critical input standard deviation $\sigma_{\text{in},0}$ at which the transition occurs gradually decreases as $N$ increases.

\subsection{Scaling for the noise intensity}

%%%%%%%%%%%%%%%%%%%%%%%%%%%%%%%%%%%%
\begin{figure}
\centering
\includegraphics[width=8.6cm]{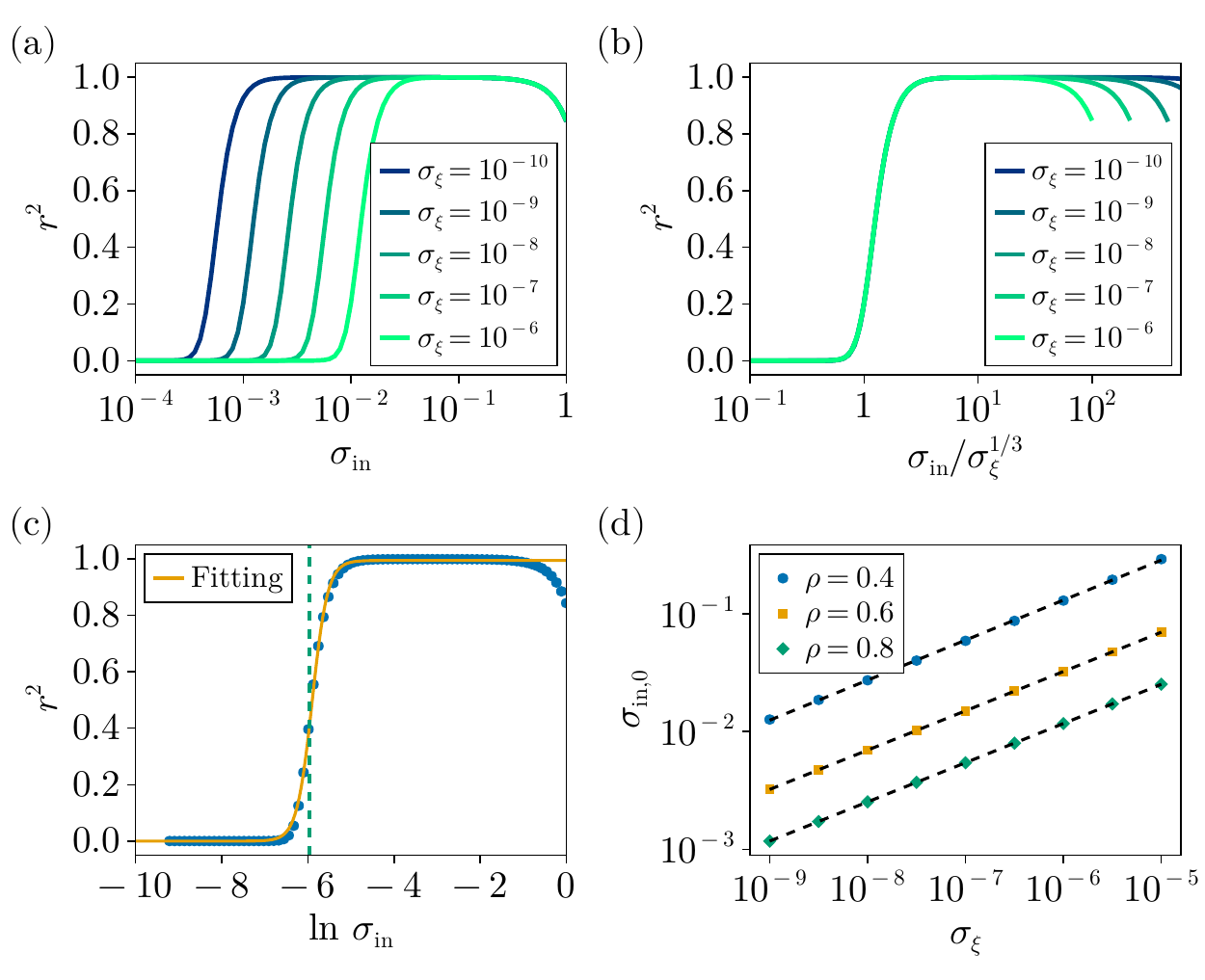}
\caption{Scaling of the transition point with respect to the noise intensity $\sigma_{\xi}$.
(a) Information processing capacity $r^2$ as a function of the input standard deviation $\sigma_{\text{in}}$ for various noise intensities $\sigma_{\xi}=10^{-10}$, $10^{-9}$, $10^{-8}$, $10^{-7}$, $10^{-6}$.
The network has $N=1000$ nodes, and the spectral radius of $W$ is $\rho=0.8$.
The XOR task uses a delay of $k=10$.
(b) $r^2$ plotted against the rescaled input standard deviation $\sigma_{\text{in}}/\sigma_{\xi}^{1/3}$.
The collapse of the curves for different $\sigma_{\xi}$ onto a single master curve is clearly observed.
(c) Example of fitting $r^2$ versus $\sigma_{\text{in}}$ for $\sigma_{\xi}=10^{-8}$ using a logarithmic logistic function.
The dashed line indicates the transition point $\sigma_{\text{in}, 0}$, defined as the point where the derivative $dr^2/d\sigma_{\text{in}}$ attains its maximum.
(d) $\sigma_{\text{in}, 0}$ as a function of the noise intensity $\sigma_{\xi}$ for different spectral radii $\rho = 0.4$, $0.6$, $0.8$.
The dashed lines show power-law fits of the form $\sigma_{\text{in}, 0} \propto \sigma_{\xi}^\eta$, with $\eta=0.340$ for $\rho=0.4$, $\eta=0.334$ for $\rho=0.6$, and $\eta=0.333$ for $\rho=0.8$.}
\label{fig_r2_noise_scaling}
\end{figure}
%%%%%%%%%%%%%%%%%%%%%%%%%%%%%%%%%%%%

Let us consider how the noise affects the capacity $r^2$.
Figure \ref{fig_r2_noise_scaling}(a) shows $r^2$ as a function of $\sigma_{\text{in}}$ for various noise intensities $\sigma_{\xi}$.
As expected, large noise diminishes the capacity.
Notably, in Fig.~\ref{fig_r2_noise_scaling}(b), when $r^2$ is plotted against the rescaled input standard derivation $\sigma_{\text{in}}/\sigma_{\xi}^{1/3}$, the curves corresponding to various $\sigma_{\xi}$ collapse onto a single master curve.
We thus expect a scaling law for the critical input standard deviation $\sigma_{\text{in},0}$ at which the linear-to-nonlinear transition occurs,
\begin{equation}
\sigma_{\text{in},0} \propto \sigma_{\xi}^{1/3}.
\label{transition_noise_scaling}
\end{equation}

To verify the scaling law in Eq.~\eqref{transition_noise_scaling}, we examine the behavior of $\sigma_{\text{in},0}$ as a function of the noise intensity $\sigma_{\xi}$.
We define the transition point $\sigma_{\text{in},0}$ as the value of $\sigma_{\text{in}}$ at which the derivative $dr^2/d\sigma_{\text{in}}$ is maximized,
\begin{equation}
\sigma_{\text{in}, 0} = \argmax_{\sigma_{\text{in}}} \frac{dr^2(\sigma_{\text{in}})}{d\sigma_{\text{in}}}. 
\label{def_sigma_in_0}
\end{equation}
To reduce the fluctuations in $\sigma_{\text{in}, 0}$, we fit $r^2$ as a function of $\sigma_{\text{in}}$ using a logarithmic logistic function of the form,
\begin{equation}
f(x) = \frac{a}{1 + e^{-b (\ln x - c)}},
\label{logarithmic_logistic_function}
\end{equation}
with fitting parameters $a$, $b$, and $c$ [see Fig.~\ref{fig_r2_noise_scaling}(c)].
We then obtain $\sigma_{\text{in}, 0}$ directly from these fitted parameters:
\begin{equation}
\sigma_{\text{in}, 0} = e^c \left( \frac{b-1}{b+1} \right)^{1/b}.
\label{sigma_in_0_fitting}
\end{equation}
Figure \ref{fig_r2_noise_scaling}(d) shows $\sigma_{\text{in}, 0}$ as a function of the noise intensity $\sigma_{\xi}$ for different spectral radii.
The dashed lines show power-law fits of the form $\sigma_{\text{in}, 0} \propto \sigma_{\xi}^\eta$, with $\eta=0.340$ for $\rho=0.4$, $\eta=0.334$ for $\rho=0.6$, and $\eta=0.333$ for $\rho=0.8$, confirming the scaling law in Eq.~\eqref{transition_noise_scaling}.

The emergence of a nontrivial exponent in the scaling relation \eqref{transition_noise_scaling} suggests that the transition from the low-capacity phase (1) to the high-capacity phase (2) is not simply a consequence of an improved signal-to-noise ratio.  
Since the magnitude of the network state $x_i(t)$ scales proportionally with the input standard deviation $\sigma_{\text{in}}$, the signal-to-noise ratio is given by $\sigma_{\text{in}} / \sigma_{\xi}$.  
If the transition between phases (1) and (2) were governed solely by the signal-to-noise ratio, the critical input standard deviation would be expected to scale as $\sigma_{\text{in},0} \propto \sigma_{\xi}$.  
However, the observed scaling relation \eqref{transition_noise_scaling} implies a weaker dependence, indicating that a distinct mechanism underlies the transition.  
In particular, it suggests the existence of an intermediate regime $\sigma_{\xi} \ll \sigma_{\text{in}} \ll \sigma_{\xi}^{1/3}$, where linear information processing remains robust and is not significantly degraded by noise.  

The scaling relation \eqref{transition_noise_scaling} implies that, in the limit of vanishing noise $\sigma_{\xi} \to 0$, $\sigma_{\text{in},0}$ goes to zero.
In other words, for a large network, even infinitesimally weak nonlinearity is sufficient to drive the system into a nonlinear information processing regime.
This observation is consistent with previous findings showing that introducing a small number of nonlinear nodes into an otherwise linear network can significantly enhance its information processing capacity \cite{Inubushi-17}.  
It also recalls the phenomenon in which adding infinitesimally weak nonlinearity to an integrable system can lead to global chaos and full thermalization in the thermodynamic limit, observed in classical anharmonic lattices and quantum spin systems \cite{Berman-05, D'Alessio-16, Langen-16}.
Nevertheless, we stress that our transition from linear to nonlinear information processing is not related to the onset of chaos.

The scaling exponent $\eta = 1/3$ might be understood from the analytic properties of the activation function $h(x)$ in Eq.~\eqref{network_evolution}.
When both $\sigma_{\text{in}}$ and $\sigma_{\xi}$ are small, the network dynamics are primarily governed by the behavior of $h(x)$ near $x = 0$.
For example, the commonly used sigmoid-type activation function $h(x) = \tanh(x)$ admits a Taylor expansion of the form $h(x) = x + O(x^3)$ around the origin.
We conjecture that the observed exponent $\eta = 1/3$ arises from the cubic term $O(x^3)$, which represents the leading-order nonlinearity beyond the linear regime.
To test this conjecture, we examine in Appendix \ref{sec:activation_function} the scaling behavior with respect to the noise intensity using alternative activation functions.
For an activation function whose Taylor expansion begins at fifth order, i.e., $h(x) = x + O(x^5)$, we find that the scaling exponent becomes $\eta = 1/5$ [see Figs.~\ref{fig_activation_noise_scaling}(c) and (d)].
In contrast, for a piecewise linear activation function that behaves exactly as $h(x)=x$ within the range $|x|<1$, the transition point $\sigma_{\text{in},0}$ remains constant regardless of the noise intensity, yielding $\eta = 0$ [see Fig.~\ref{fig_activation_noise_scaling}(e)].
These results support the conjecture that the scaling exponent $\eta$ is determined by the inverse of the order of the lowest-order nonlinear term in the expansion of $h(x)$ around $x = 0$.

It is worth noting that noise has a similar effect to the $L^2$ regularization used in ridge regression.  
This form of regularization can be introduced by minimizing the following loss function:
\begin{equation}
E_{\text{ridge}}[\{w_i^{\text{out}}\}] := \sum_{t=1}^T |y(t) - z(t)|^2 + \beta \sum_{i=0}^N |w_i^{\text{out}}|^2,
\end{equation}
where $\beta$ is the ridge parameter.  
As discussed in Appendix~\ref{sec:ridge_regression}, $\beta$ corresponds to an effective noise standard deviation $\sigma_{\xi} \sim \beta^{1/2}$.  
Consequently, in the ridge regression case, the appropriate rescaling of the input standard deviation is given by $\sigma_{\text{in}} / \beta^{1/6}$, as demonstrated in Fig.~\ref{fig_r2_ridge_noise_scaling}.

\subsection{Scaling for the network size}

%%%%%%%%%%%%%%%%%%%%%%%%%%%%%%%%%%%%
\begin{figure}
\centering
\includegraphics[width=8.6cm]{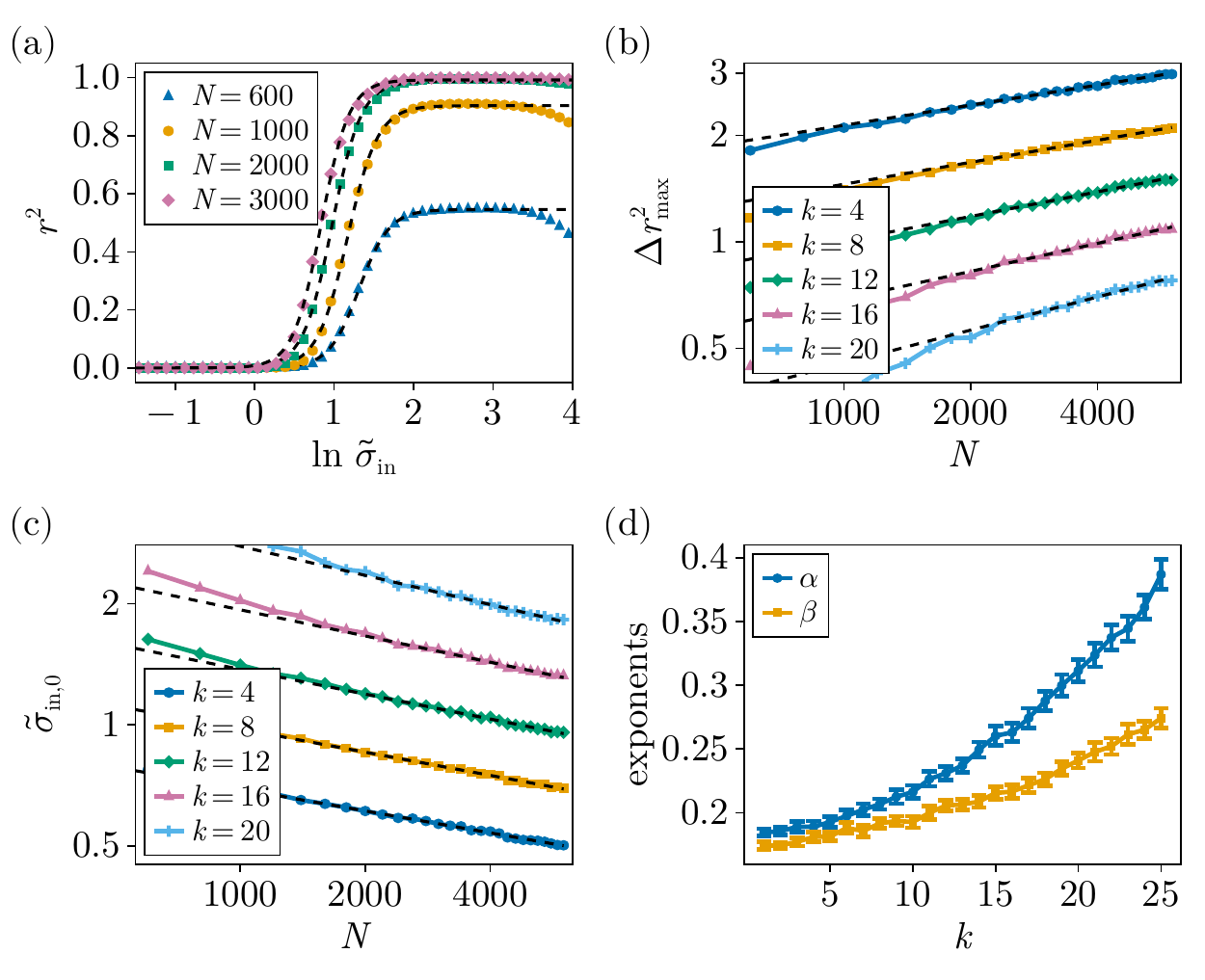}
\caption{Network size scaling of the transition point.
(a) Information processing capacity $r^2$ versus the rescaled input standard deviation $\tilde{\sigma}_{\text{in}}=\sigma_{\text{in}}/\sigma_{\xi}^{1/3}$ for $\sigma_{\xi}=10^{-8}$ at several network sizes $N$ (XOR task with delay $k=20$).
The dashed curves show logarithmic-logistic fits.
The peak value of $r^2$ saturates to $1$ for $N \gtrsim 2000$, beyond which both $\Delta r^2_{\text{max}}$ and $\tilde{\sigma}_{\text{in},0}$ exhibit power-law scaling with respect to $N$.
(b) Network size dependence of the maximum derivative $\Delta r^2_{\text{max}}$ at the transition point in a double-log plot.
The dashed lines represent $\Delta r^2_{\text{max}} \propto N^{\alpha}$.
(c) Network size dependence of the transition point $\tilde{\sigma}_{\text{in}, 0}$ in a double-log plot.
The dashed lines represent $\tilde{\sigma}_{\text{in}, 0} \propto N^{-\beta}$.
(d) Exponents $\alpha$ and $\beta$ for different delay values $k$.
The power-law fits for $\tilde{\sigma}_{\text{in}, 0}$ and $\Delta r^2_{\text{max}}$ are performed for $N \geq 2200$.
The error bars indicate the standard errors from the fitting.
The fact that $\alpha > \beta$ implies that the transition sharpens faster than $\tilde{\sigma}_{\text{in}, 0}$ vanishes.}
\label{fig_r2_size_scaling}
\end{figure}
%%%%%%%%%%%%%%%%%%%%%%%%%%%%%%%%%%%%

We next consider how increasing the network size $N$ influences the location and sharpness of the transition.
In the following, we consider the information processing capacity $r^2$ as a function of the rescaled input standard deviation 
\begin{equation}
\tilde{\sigma}_{\text{in}} := \sigma_{\text{in}}/\sigma_{\xi}^{1/3}.
\end{equation}

As in Eq.~\eqref{def_sigma_in_0}, we define the rescaled critical point $\tilde{\sigma}_{\text{in}, 0}$ as the value of $\tilde{\sigma}_{\text{in}}$ at which the derivative $dr^2/d\tilde{\sigma}_{\text{in}}$ is maximized.
In addition, we denote the  maximum of the derivative by $\Delta r^2_{\text{max}}$:
\begin{equation}
\Delta r^2_{\text{max}} = \frac{dr^2(\tilde{\sigma}_{\text{in}, 0})}{d\tilde{\sigma}_{\text{in}}}.
\end{equation}
The maximum derivative $\Delta r^2_{\text{max}}$ represents the sharpness of the transition.

To reduce the fluctuations in $\tilde{\sigma}_{\text{in}, 0}$ and $\Delta r^2_{\text{max}}$, we fit $r^2$ as a function of $\tilde{\sigma}_{\text{in}}$ using a logarithmic logistic function given by Eq.~\eqref{logarithmic_logistic_function} with fitting parameters $a$, $b$, and $c$.
In addition to Eq.~\eqref{sigma_in_0_fitting}, $\Delta r^2_{\text{max}}$ is given by
\begin{equation}
\Delta r^2_{\text{max}} = e^{-c} \frac{a (b^2 - 1)}{4b} \left( \frac{b+1}{b-1} \right)^{1/b}.
\end{equation}

Figure \ref{fig_r2_size_scaling}(a) presents examples of fits of $r^2$ versus the rescaled input standard deviation $\tilde{\sigma}_{\text{in}}$ for the XOR task with delay $k=20$.
The noise intensity is $\sigma_\xi = 10^{-8}$, and the spectral radius of $W$ is $\rho=0.8$.
The results show that the peak value of $r^2$ saturates to unity for $N \gtrsim 2000$.
Figure \ref{fig_r2_size_scaling}(b) shows the size dependence of $\Delta r^2_{\text{max}}$ in a double-log plot for different delay values $k$ in the XOR task.
For sufficiently large $N$, we find that the sharpness of the transition scales as $\Delta r^2_{\text{max}} \propto N^{\alpha}$ with a positive exponent $\alpha$.
Figure \ref{fig_r2_size_scaling}(c) shows the location of the transition $\tilde{\sigma}_{\text{in}, 0}$ as a function of $N$, which scales as $\tilde{\sigma}_{\text{in}, 0} \propto N^{-\beta}$ with a positive exponent $\beta$.
Note that both $\Delta r^2_{\text{max}}$ and $\tilde{\sigma}_{\text{in},0}$ enter the power-law regime once the peak value of $r^2$ versus $\tilde{\sigma}_{\text{in}}$ saturates to unity.

Figure \ref{fig_r2_size_scaling}(d) shows the exponents $\alpha$ and $\beta$ as functions of $k$.
The power-law fits for $\tilde{\sigma}_{\text{in}, 0}$ and $\Delta r^2_{\text{max}}$ are performed over the range $N \in [2200, 6000]$.
As $k$ increases, the XOR task demands longer memory, making it more difficult.
We observe that both $\alpha$ and $\beta$ grow with $k$, although we cannot rule out the possibility that this trend is influenced by the choice of fitting range.
Appendix \ref{sec:finite_size_scaling_suppl} provides a detailed discussion of how the fitting range affects the extracted values of $\alpha$ and $\beta$.
We find that $\alpha > \beta$, which implies that the sharpness of the transition increases more rapidly than the decrease in $\tilde{\sigma}_{\text{in}, 0}$.
This result suggests that, when plotting $r^2$ against $\tilde{\sigma}_{\text{in}}N^\beta$, a discontinuous transition may emerge in the limit $N \to \infty$, as illustrated in Fig.~\ref{fig_transition_schematic}.

\subsection{Other tasks}

To demonstrate the robustness of the transition discussed above, we evaluate the information processing capacity on three additional benchmark problems:
\begin{enumerate}
\item \textbf{Nonlinear autoregressive moving average (NARMA) task}.
The target is generated by 
\begin{align}
z(t+1) =& \ a_1 z(t) + a_2 z(t) \sum_{i=0}^{m-1} z(t-i) \nonumber \\
&+ a_3 u(t-m+1) u(t) + a_4,
\end{align}
with parameters $m=10$, $a_1=0.3$, $a_2=0.05$, $a_3=0.2$, and $a_4=0.05$ \cite{Rodan-11, Appeltant-11}.
The input is a random binary sequence $u(t) \in \{0,1\}$, as in the XOR task.

\item \textbf{One-step prediction task of the Lorenz system}.
We consider the Lorenz system, a canonical example of chaotic dynamics, defined by
\begin{align}
\frac{dX}{dt} &= \sigma (Y - X), \nonumber \\
\frac{dY}{dt} &= X(\rho - Z) - Y, \\
\frac{dZ}{dt} &= X Y - \beta Z, \nonumber
\end{align}
with parameters $\rho=28$, $\sigma=10$, and $\beta=8/3$.
The input sequence $u(t)$ is generated by sampling $X(t)$ at intervals of $0.2$, adding Gaussian noise with standard deviation $0.1$ to simulate observation noise, and then normalizing to zero mean and unit variance.  
The target is set to $z(t) = u(t+1)$, corresponding to a one-step-ahead prediction task based solely on the time series of $X$, rather than the full state $(X, Y, Z)$.
This setup reflects realistic scenarios of predicting partially observed chaotic time series.

\item \textbf{Delay task}. 
The target is defined as $z(t)=u(t-k)$, which simply requires the network to recall the input with a delay of $k$.
We again use a random binary sequence $u(t) \in \{0, 1\}$.
\end{enumerate}

%%%%%%%%%%%%%%%%%%%%%%%%%%%%%%%%%%%%
\begin{figure}
\centering
\includegraphics[width=8.6cm]{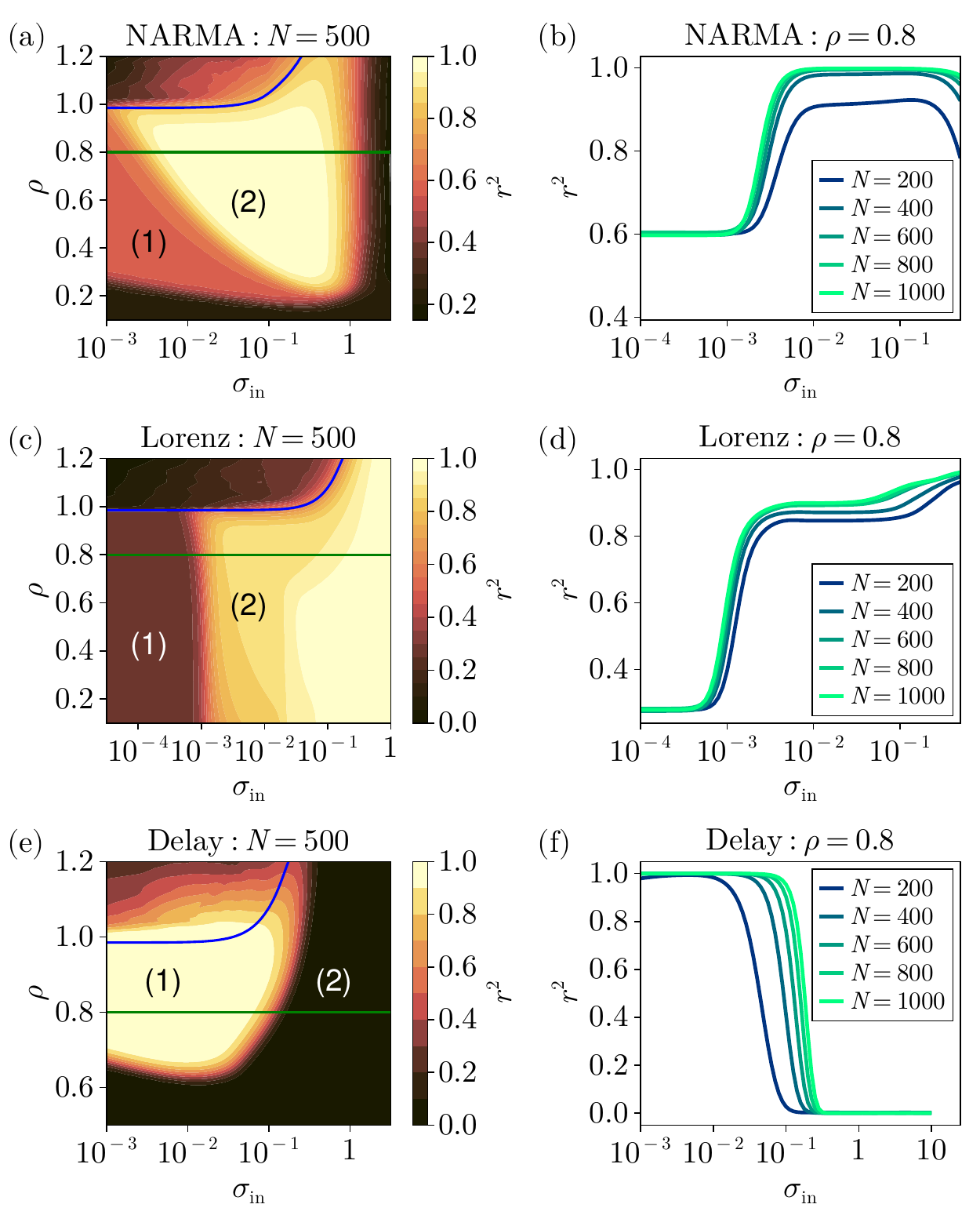}
\caption{Information processing capacity $r^2$ for the NARMA, Lorenz, and delay tasks.
(a) $r^2$ for the NARMA task as a function of $\sigma_{\text{in}}$ and $\rho$ with $N=500$.
(b) $r^2$ for the NARMA task with $\rho=0.8$ and various values of $N$.
(c) $r^2$ for the Lorenz task as a function of $\sigma_{\text{in}}$ and $\rho$ with $N=500$.
(d) $r^2$ for the Lorenz task with $\rho=0.8$ and various values of $N$.
(e) $r^2$ for the delay task as a function of $\sigma_{\text{in}}$ and $\rho$ with $N=500$.
The value of delay is $k=30$.
(f) $r^2$ for the delay task with $\rho=0.8$ and various values of $N$.
The regions labeled (1) and (2) indicate the linear and nonlinear information processing phases, respectively.
Panels (b), (d), and (f) show that the boundary between phases (1) and (2) becomes shaper as $N$ increases.}
\label{fig_r2_narma_delay}
\end{figure}
%%%%%%%%%%%%%%%%%%%%%%%%%%%%%%%%%%%%

Figures \ref{fig_r2_narma_delay}(a) and (b) show the capacity $r^2$ for the NARMA task.
The blue solid curve in (a) indicates the boundary of the order-to-chaos transition.
The regions labeled (1) and (2) correspond to the linear and nonlinear information processing phases, respectively.
In contrast to the XOR task, the NARMA task yields a nonzero $r^2$ even within the linear regime.
Nevertheless, a sharp transition from phase (1) to phase (2) is still clearly observed.
Furthermore, as discussed in Appendix \ref{sec:narma_delay}, when $r^2$ is plotted against $\sigma_{\text{in}}/\sigma_{\xi}^{1/3}$, the curves for different noise levels collapse onto a single curve, confirming the universality of the scaling behavior Eq.~\eqref{transition_noise_scaling}.

It is worth noting that, for the NARMA task, the linear regime (1), in which the capacity $r^2$ is nonzero, does not persist down to arbitrarily small values of $\sigma_{\text{in}}$.  
When $\sigma_{\text{in}}$ becomes comparable to the noise intensity $\sigma_{\xi}$, the signal-to-noise ratio deteriorates, resulting in a significant drop in information processing capacity.
Given that the critical input standard deviation scales according to Eq.~\eqref{transition_noise_scaling}, robust linear information processing is effectively realized within the intermediate regime $\sigma_{\xi} \ll \sigma_{\text{in}} \ll \sigma_{\xi}^{1/3}$.  

Figures \ref{fig_r2_narma_delay}(c) and (d) show the capacity $r^2$ for the one-step prediction task of the Lorenz system.
As in the NARMA task, $r^2$ remains nonzero even within the linear regime, indicated by label (1).
In contrast to the XOR and NARMA tasks, the nonlinear regime [label (2)] exhibits a two-step structure, featuring an intermediate plateau in $r^2$.
Despite this difference, a clear boundary between the linear and nonlinear regimes can be identified, and it follows the same noise scaling as observed for the XOR and NARMA tasks (see Appendix~\ref{sec:narma_delay}).  
Note that the location of the boundary between regimes (1) and (2) is nearly independent of the spectral radius $\rho$ of the recurrent weight matrix.
In general, the spectral radius $\rho$ governs the characteristic time scale of the network dynamics; as $\rho \to 1$, the relaxation time of the undriven network diverges.
The nearly vertical phase boundary observed in Fig.~\ref{fig_r2_narma_delay}(c) therefore suggests that the Lorenz prediction task does not require long-term memory.

Figures \ref{fig_r2_narma_delay}(e) and (f) show the capacity $r^2$ for the delay task with a delay parameter $k=30$.
It is well known that there exists a trade-off between nonlinearity and memory capacity \cite{Dambre-12, Inubushi-17}, the latter quantifying how well the system retains information about past inputs.
Since the delay task relies purely on memory, linear networks tend to achieve higher $r^2$ values than their nonlinear counterparts.
As a result, $r^2$ drops to zero when the system transitions from the linear to the nonlinear regime.
In contrast to the XOR and NARMA tasks, we find that the critical value of the {\it unscaled} input standard deviation $\sigma_{\text{in}}$ at which the transition occurs is independent of the noise strength $\sigma_{\xi}$ (see Appendix \ref{sec:narma_delay}).
This suggests that the transition in the delay task may belong to a different universality class than those observed in the XOR and NARMA tasks.

\section{Conclusion}
\label{sec:conclusion}

In this study, we investigated the phase transition from linear to nonlinear information processing in neural networks. 
Our results suggest that the information processing capacity exhibits a discontinuous jump in the limit of infinitely many nodes when plotted against an appropriately rescaled nonlinearity parameter.
This transition is distinct from conventional order-chaos bifurcations and equilibrium thermodynamic phase transitions, and thus may enlarge the catalog of nonequilibrium phase transitions in high-dimensional dynamical systems.

At present, we do not have a theoretical explanation for the phase transition reported in this study.
A commonly used analytical approach for recurrent neural networks with random weights is the dynamical mean-field theory \cite{Sompolinsky-88, Kadmon-15, Schuecker-18}.
For instance, Ref.~\cite{Schuecker-18} applied this approach to compute the information processing capacity for the delay task in a continuous-time random neural network.
However, to the best of our knowledge, no singular behavior of the capacity, as observed in our work, has been reported within this framework.
Developing a theoretical understanding of this transition remains an important direction for future research.

In our study, the nonlinearity of the network is controlled by the standard deviation of the input weights.
Alternatively, one can consider introducing nonlinearity by incorporating a small number of nonlinear nodes into an otherwise linear network.
Our results suggest that even a small fraction of nonlinear nodes can substantially enhance the computational capability of such a network, consistent with the qualitative findings reported in Ref.~\cite{Inubushi-17}.
An intriguing open question is whether there exists a sharp transition at a critical fraction $f_c$ of nonlinear nodes, below which the network effectively behaves as linear and above which it achieves an information processing capacity comparable to a fully nonlinear network.
Investigating this possibility can provide valuable guidance for identifying the optimal balance between linear and nonlinear nodes in the design of reservoir computing systems.

Although our analysis has focused on neural networks, the type of transition described in this study is likely to emerge in a broad class of physical systems.  
As discussed in Sec.~\ref{sec:information_processing_capacity}, the concept of information processing capacity is general and can be applied to a wide range of dynamical systems, including quantum systems.  
Such systems often possess control parameters that regulate their degree of nonlinearity, for example, anharmonic terms in coupled harmonic oscillators or next-nearest-neighbor couplings in quantum spin chains.  
This naturally raises the question of whether the transition from linear to nonlinear information processing observed in neural networks also occurs in these physically motivated systems.  

A low noise level is crucial for observing the transition between linear and nonlinear information processing regimes in physical implementations of reservoir computing.
Note that the transition occurs when the rescaled input strength $\tilde{\sigma}_{\text{in}, 0} = \sigma_{\text{in}, 0} / \sigma_\xi^{1/3}$ is of order unity.
Consequently, to achieve a given transition point $\sigma_{\text{in}, 0}$ requires the noise intensity to scale as $\sigma_\xi \sim \sigma_{\text{in}, 0}^3$.
In the linear regime, the network state must remain small ($|x_n| \ll 1$), so that the activation function is effectively linear ($h(x) \simeq x$).
Since the state amplitude is proportional to the input strength $\sigma_{\text{in}}$, the critical input strength must satisfy $\sigma_{\text{in},0} \ll 1$.
Consequently, the signal-to-noise ratio required to observe the transition can be estimated as $\sigma_{\text{in},0}/\sigma_\xi \sim \sigma_{\text{in},0}^{-2}$, which is typically on the order of $10^2$ or higher.

\begin{acknowledgments}
This work was supported by JSPS KAKENHI Grant Number JP22K13983.
\end{acknowledgments}

\section*{DATA AVAILABILITY}
The data that support the findings of this article are openly available \cite{data}.

\appendix

\section{Training procedure of echo state networks}
\label{sec:training_procedure}

In this appendix, we describe the procedure for training the echo state network.
Suppose we have a sequence of network states $\mathbf{x}(1), \ldots, \mathbf{x}(T)$ and corresponding targets $z(1), \ldots, z(T)$.
The model output $y(t)$ is given by Eq.~\eqref{model_output}.
Let us define the matrix $X \in \mathbb{R}^{(N+1) \times T}$ by setting $X_{1j}=1$ and $X_{ij} = x_{i-1}(j)$ for $i \geq 2$.
The optimized weights $W^{\text{out}} = (w^{\text{out}}_0, w^{\text{out}}_1, \ldots, w^{\text{out}}_N) \in \mathbb{R}^{1 \times (N+1)}$ are obtained by
\begin{equation}
W^{\text{out}} = Z X^{\text{T}} (X X^{\text{T}})^{-1} = Z X^+,
\label{linear_regression}
\end{equation}
where $Z = (z(1), \ldots, z(T)) \in \mathbb{R}^{1 \times T}$ is the row vector of target values, and $X^+$ is the Moore-Penrose pseudoinverse of $X$.
In a \texttt{Julia} implementation of Eq.~\eqref{linear_regression}, directly using \texttt{inv} to invert $X X^{\text{T}}$ can cause numerical instability near the transition from linear to nonlinear information processing.
To avoid this issue, we compute $X^+$ via singular value decomposition, as provided by the \texttt{pinv} function.

The training time \(T\) should be larger than the number of nodes \(N\) to avoid overfitting.  
We set \(T = 5N\) and introduce a warm-up period of 200 time steps to discard the initial transient dynamics of the network.  
During the testing phase, the network is run for \(T = 2N\) time steps, and the model output \(y(t)\) is computed using Eq.~\eqref{model_output} with the trained weights \(w^{\text{out}}_i\).  
The information processing capacity is then evaluated as the square of the Pearson correlation coefficient between the model output \(y(t)\) and the target \(z(t)\).

\section{Details of network size scaling}
\label{sec:finite_size_scaling_suppl}

%%%%%%%%%%%%%%%%%%%%%%%%%%%%%%%%%%%%
\begin{figure}
\centering
\includegraphics[width=8.6cm]{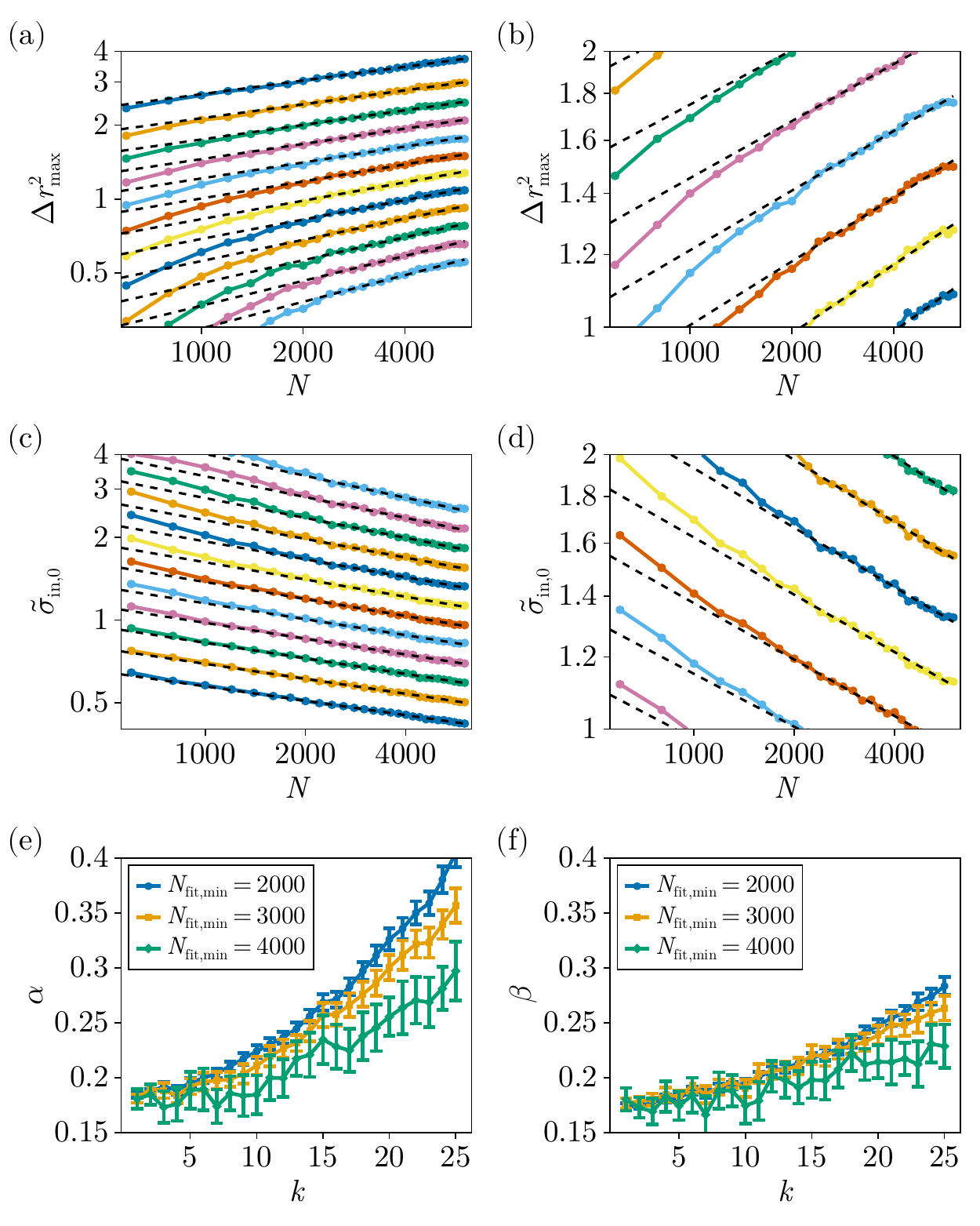}
\caption{(a) Network size dependence of the maximum derivative $\Delta r^2_{\text{max}}$ at the transition point in a double-log plot.
The values of the XOR delay are $k = 2, 4, 6, \ldots, 24$ from top to bottom.
The dashed lines represent $\Delta r^2_{\text{max}} \propto N^{\alpha}$.
(b) Enlarged view of panel (a), highlighting the power-law dependence.
(c) Network size dependence of the transition point $\tilde{\sigma}_{\text{in}, 0}$ in a double-log plot.
The values of the XOR delay are $k = 2, 4, 6, \ldots, 24$ from bottom to top.
The dashed lines represent $\tilde{\sigma}_{\text{in}, 0} \propto N^{-\beta}$.
(d) Enlarged view of panel (c).
(e), (f) Exponents $\alpha$ and $\beta$ as functions of the delay $k$, obtained using different fitting ranges.
The power-law fits for $\tilde{\sigma}_{\text{in}, 0}$ and $\Delta r^2_{\text{max}}$ are performed for $N \in [N_{\text{fit,min}}, 6000]$, with $N_{\text{fit,min}}=2000$, $3000$, and $4000$.
The error bars indicate the standard errors from the fitting.}
\label{fig_r2_size_scaling_suppl}
\end{figure}
%%%%%%%%%%%%%%%%%%%%%%%%%%%%%%%%%%%%

In this appendix, we provide a detailed analysis of how the transition depends on the network size.
We focus on two key quantities: the rescaled transition point $\tilde{\sigma}_{\text{in}, 0}$, defined as the value of $\tilde{\sigma}_{\text{in}}$ at which the derivative $dr^2/d\tilde{\sigma}_{\text{in}}$ attains its maximum, and the maximum derivative itself, denoted $\Delta r^2_{\text{max}}$.
Our aim is to examine how both $\tilde{\sigma}_{\text{in}, 0}$ and $\Delta r^2_{\text{max}}$ scale with the network size $N$.

Figure \ref{fig_r2_size_scaling_suppl}(a) shows $\Delta r^2_{\text{max}}$ as a function of $N$ for the various values of the XOR delay $k = 2, 4, 6, \ldots, 24$ from top to bottom.
The noise intensity is $\sigma_\xi = 10^{-8}$, and the spectral radius of $W$ is $\rho=0.8$.
An enlarged view is provided in Fig.~\ref{fig_r2_size_scaling_suppl}(b) to emphasize the power-law scaling behavior.
Figure \ref{fig_r2_size_scaling_suppl}(c) shows $\tilde{\sigma}_{\text{in}, 0}$ as a function of $N$ for the various values of the XOR delay $k = 2, 4, 6, \ldots, 24$ from bottom to top.
The corresponding magnified view is shown in Fig.~\ref{fig_r2_size_scaling_suppl}(d).
For sufficiently large $N$, both $\Delta r^2_{\text{max}}$ and $\tilde{\sigma}_{\text{in}, 0}$ exhibit power-law scaling as $\Delta r^2_{\text{max}} \propto N^{\alpha}$ and $\tilde{\sigma}_{\text{in}, 0} \propto N^{-\beta}$ with positive exponents $\alpha$ and $\beta$.

The key questions are: (i) how the scaling exponents $\alpha$ and $\beta$ depend on the XOR delay $k$, and (ii) whether $\alpha$ is larger than, equal to, or smaller than $\beta$.
The first question concerns the universality of the scaling exponents.
If $\alpha$ and $\beta$ are independent of the XOR delay, this would suggest that they are intrinsic properties of the system, rather than those of the specific benchmark task used to evaluate information processing capacity.
The second question addresses the emergence of singular behavior in the limit of infinite system size.
If $\alpha=\beta$, then the location and sharpness of the transition scale identically, suggesting a scaling relation of the form $r^2(\tilde{\sigma}_{\text{in}}, N) = f(\tilde{\sigma}_{\text{in}} N^\alpha)$ with a smooth function $f(x)$ connecting the linear and nonlinear capacities.
By contrast, if $\alpha > \beta$, the transition sharpens more rapidly than the transition point shifts.
This would imply that, when plotting $r^2$ against $\tilde{\sigma}_{\text{in}}N^\beta$, a discontinuous transition may emerge in the limit $N \to \infty$.

Reaching definitive conclusions on the above questions is challenging because the minimal network size $N$ required for $\Delta r^2_{\text{max}}$ and $\tilde{\sigma}_{\text{in}, 0}$ to exhibit clear power-law behavior increases with the XOR delay $k$.
As a result, for a fixed fitting range $N \in [N_{\text{fit,min}}, N_{\text{fit,max}}]$, the estimated exponents $\alpha$ and $\beta$ may display an artificial $k$-dependence due to transient effects, even if they are in fact constant.
Figures \ref{fig_r2_size_scaling_suppl}(e) and (f) show $\alpha$ and $\beta$ as functions of $k$ for fitting ranges $[N_{\text{fit,min}}, 6000]$ with $N_{\text{fit,min}} = 2000$, $3000$, and $4000$.
A naive interpretation suggests that both $\alpha$ and $\beta$ increase with $k$, and $\alpha > \beta$.
However, as $N_{\text{fit,min}}$ increases, the values of $\alpha$ and $\beta$ decrease for large $k$, resulting in a weaker $k$-dependence.
If this trend continues at larger $N$, it is possible that both exponents converge to a common universal value of approximately $0.18$, independent of $k$.
Given the limitations of numerically accessible system sizes, we cannot rule out this possibility.
Further investigations with larger networks will be required to settle this question conclusively.

\section{Ridge regression}
\label{sec:ridge_regression}

%%%%%%%%%%%%%%%%%%%%%%%%%%%%%%%%%%%%
\begin{figure}
\centering
\includegraphics[width=8.6cm]{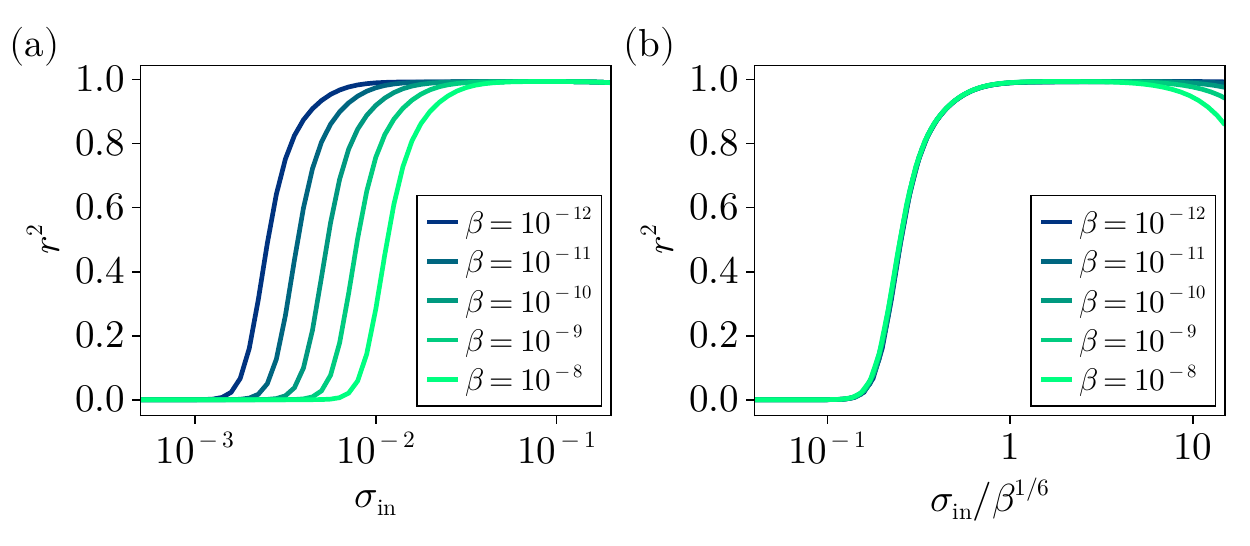}
\caption{Information processing capacity $r^2$ as a function of $\sigma_{\text{in}}$ for various ridge parameters $\beta=10^{-12}, 10^{-11}, 10^{-10}, 10^{-9}, 10^{-8}$.
The network has $N=1000$ nodes, and the spectral radius of $W$ is $\rho=0.8$.
The noise intensity $\sigma_{\xi}$ is zero.
The XOR task uses a delay of $k=10$.
In panel (b), $r^2$ is plotted against $\sigma_{\text{in}}/\beta^{1/6}$.
The collapse of the curves for different $\beta$ onto a single master curve is clearly observed.}
\label{fig_r2_ridge_noise_scaling}
\end{figure}
%%%%%%%%%%%%%%%%%%%%%%%%%%%%%%%%%%%%

In this appendix, we discuss how the main results are affected when ridge regression is used in place of standard linear regression.
Using the same notation as in Eq.~\eqref{linear_regression}, the optimized weights are given by
\begin{equation}
W^{\text{out}} = Z X^{\text{T}} (X X^{\text{T}} + \beta I)^{-1},
\label{ridge_regression}
\end{equation}
where $\beta$ is the ridge parameter and $I$ is the $(N+1) \times (N+1)$ identity matrix.

The effect of the ridge parameter $\beta$ can be understood as analogous to the introduction of a noise.
In fact, if a mean-zero independent noise with variance $\sigma^2$ is added to the network state matrix $X$, then $X X^{\text{T}}$ is effectively replaced by $X X^{\text{T}} + T \sigma^2 I$, where $T$ is the length of the time series.  
Although in our model the noise $\boldsymbol{\xi}(t)$ introduced in Eq.~\eqref{network_evolution} is not strictly additive due to the nonlinear activation $h(x)$, the ridge term $\beta$ is expected to have a similar regularizing effect.  
In particular, we can interpret $\beta$ as corresponding to an effective noise standard deviation $\sigma_{\xi} \sim \beta^{1/2}$.  

This expectation is confirmed in Fig.~\ref{fig_r2_ridge_noise_scaling}, which shows the information processing capacity \(r^2\) as a function of \(\sigma_{\text{in}}\) for various values of the ridge parameter \(\beta\).  
Here, the noise intensity is set to zero, i.e., \(\sigma_{\xi} = 0\).  
We observe that the curves corresponding to different \(\beta\) values collapse onto a single master curve when the input standard deviation is rescaled as \(\sigma_{\text{in}} / \beta^{1/6}\).  
This result should be compared with the case of standard linear regression, where the appropriate rescaling is \(\sigma_{\text{in}} / \sigma_{\xi}^{1/3}\).  
The consistency between the two scalings supports the correspondence \(\sigma_{\xi} \sim \beta^{1/2}\) between noise and ridge regularization.

\section{Noise scaling for other tasks}
\label{sec:narma_delay}

%%%%%%%%%%%%%%%%%%%%%%%%%%%%%%%%%%%%
\begin{figure}
\centering
\includegraphics[width=8.6cm]{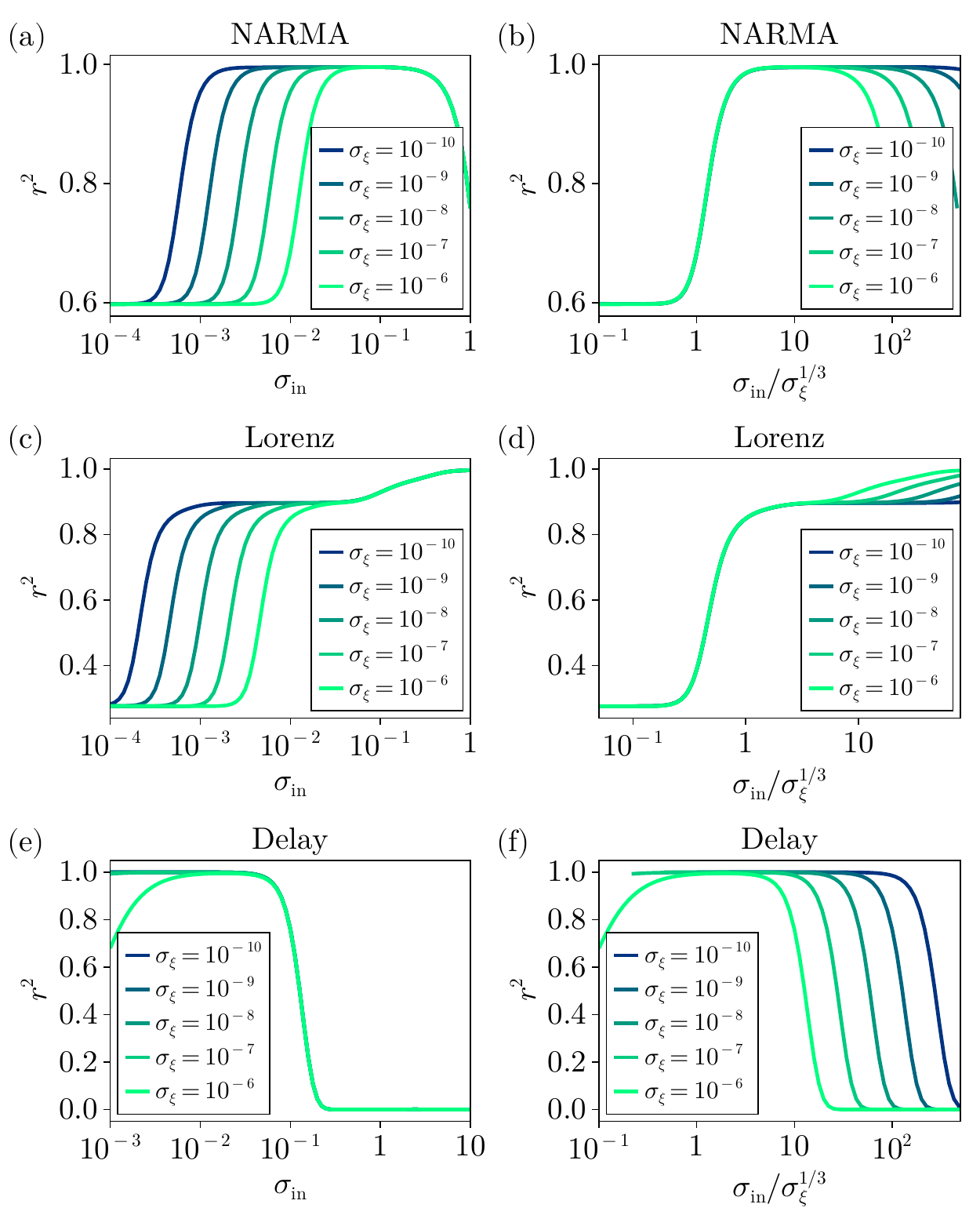}
\caption{Information processing capacity \(r^2\) as a function of \(\sigma_{\text{in}}\) for the NARMA task [panels (a) and (b)], the Lorenz task [panels (c) and (d)], and the delay task [panels (e) and (f)].  
The number of nodes is \(N = 600\), and the spectral radius of \(W\) is set to \(\rho = 0.8\).  
The delay task is performed with a delay of \(k = 30\).  
For the NARMA and Lorenz tasks, the curves corresponding to different noise intensities collapse onto a single master curve when \(\sigma_{\text{in}}\) is rescaled as \(\sigma_{\text{in}} / \sigma_{\xi}^{1/3}\).  
In contrast, for the delay task, a similar collapse is observed without any rescaling of \(\sigma_{\text{in}}\).}
\label{fig_narma_delay_noise_scaling}
\end{figure}
%%%%%%%%%%%%%%%%%%%%%%%%%%%%%%%%%%%%

In this appendix, we discuss how noise influences the information processing capacity $r^2$ for the NARMA, Lorenz, and delay tasks.
Figure~\ref{fig_narma_delay_noise_scaling}(a) shows $r^2$ as a function of the input standard deviation $\sigma_{\text{in}}$ for the NARMA task.  
A similar trend to that observed in the XOR task [see Fig.~\ref{fig_r2_noise_scaling}(a)] is evident, with the exception that $r^2$ remains nonzero in the linear regime due to the partially linear nature of the NARMA task.  
In the linear regime, $r^2 \simeq 0.6$ and is nearly independent of the noise intensity $\sigma_{\xi}$, indicating that linear information processing is robust against noise and that the transition to the nonlinear regime is not driven by an improved signal-to-noise ratio.
Figure~\ref{fig_narma_delay_noise_scaling}(b) presents $r^2$ as a function of the rescaled input standard deviation $\sigma_{\text{in}} / \sigma_{\xi}^{1/3}$.  
The collapse of the curves corresponding to different noise levels indicates the same scaling behavior observed for the XOR task.  
The same behavior can also be observed for the Lorenz task, as shown in Figs.~\ref{fig_narma_delay_noise_scaling}(c) and (d).

Figure~\ref{fig_narma_delay_noise_scaling}(e) shows $r^2$ as a function of the input standard deviation $\sigma_{\text{in}}$ for the delay task.
In this case, $r^2$ drops to zero as the system transitions from the linear to the nonlinear information processing regime.  
This behavior arises because nonlinearity degrades the memory capacity of the network, which is essential for solving the delay task.
For a relatively large noise level $\sigma_{\xi} = 10^{-6}$, we observe a reduction in $r^2$ at small values of $\sigma_{\text{in}}$.  
This reduction is attributed to the detrimental effect of noise on the signal-to-noise ratio when the input amplitude is low.
Importantly, the critical value of $\sigma_{\text{in}}$ at which the transition occurs appears to be independent of the noise level.  
This observation suggests that, in the case of the delay task, the transition is not influenced by the $L^2$ regularization typically introduced via ridge regression.

\section{Influence of activation function}
\label{sec:activation_function}

%%%%%%%%%%%%%%%%%%%%%%%%%%%%%%%%%%%%
\begin{figure}
\centering
\includegraphics[width=8.6cm]{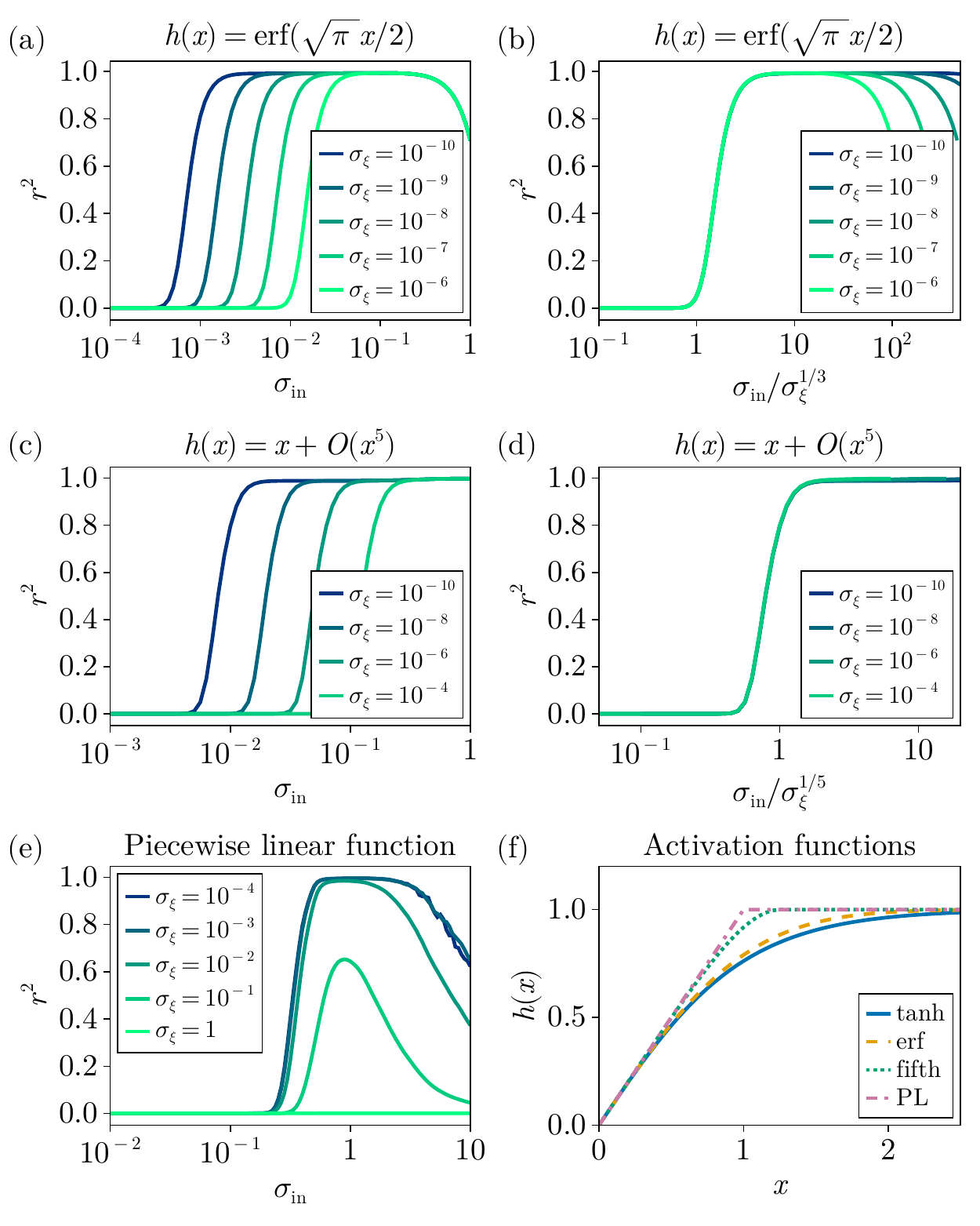}
\caption{Information processing capacity \(r^2\) as a function of \(\sigma_{\text{in}}\) for the XOR task with different activation functions $h(x)$.
Panels (a) and (b) correspond to the error function $h(x) = \text{erf}(\sqrt{\pi}x/2)$, (c) and (d) to a fifth-order function $h(x) = x + O(x^5)$, and (e) to a piecewise linear function.
The corresponding activation functions are plotted in panel (f).
The number of nodes is \(N = 600\), and the spectral radius of \(W\) is set to \(\rho = 0.8\).  
The delay in the XOR task is $k = 10$ for panels (a) and (b), and $k = 5$ for panels (c)-(e).
For the error function [panels (a) and (b)], the curves for different noise intensities $\sigma_{\xi}$ collapse onto a single master curve when $\sigma_{\text{in}}$ is rescaled as $\sigma_{\text{in}} / \sigma_{\xi}^{1/3}$.
For the fifth-order function [panels (c) and (d)], the appropriate rescaling becomes $\sigma_{\text{in}} / \sigma_{\xi}^{1/5}$, reflecting the order of the nonlinearity in $h(x)$.
For the piecewise linear function [panel (e)], the capacity $r^2$ remains essentially independent of the noise intensity when it is sufficiently small.}
\label{fig_activation_noise_scaling}
\end{figure}
%%%%%%%%%%%%%%%%%%%%%%%%%%%%%%%%%%%%

In this appendix, we examine how the choice of activation function $h(x)$ in Eq.~\eqref{network_evolution} influences the noise scaling behavior associated with the transition from linear to nonlinear information processing. 
We consider the following three types of activation functions:
\begin{enumerate}
\item \textbf{Error function}. 
\begin{equation}
h(x) = \text{erf}\left( \frac{\sqrt{\pi}}{2} x \right), \quad \text{erf}(x) := \frac{2}{\sqrt{\pi}} \int_0^x e^{-t^2} dt.
\end{equation}

\item \textbf{Fifth-order function}. 
\begin{equation}
h(x) = 
\begin{cases}
-1 &(x < -5/4), \\
x - \frac{1}{5} \left(\frac{4}{5}\right)^4 x^5 &(-5/4 \leq x < 5/4), \\
1 &(x \geq 5/4).
\end{cases}
\end{equation}

\item \textbf{Piecewise linear function}. 
\begin{equation}
h(x) = 
\begin{cases}
-1 &(x < -1), \\
x &(-1 \leq x < 1), \\
1 &(x \geq 1).
\end{cases}
\end{equation}

\end{enumerate}
Each activation function admits an expansion of the form $h(x) = x + O(x^\zeta)$ around $x=0$, where $\zeta$ denotes the order of the leading nonlinear correction. 
Specifically, we have $\zeta=3$ for the error function, $\zeta=5$ for the fifth-order function, and $\zeta=\infty$ for the piecewise linear function.
The activation functions are shown in Fig.~\ref{fig_activation_noise_scaling}(f).
In the remainder of this section, we investigate the relationship between $\zeta$ and the scaling exponent $\eta$, which characterizes how the transition point $\sigma_{\text{in}, 0}$ scales with the noise intensity $\sigma_{\xi}$.

Figure \ref{fig_activation_noise_scaling}(a) shows the information processing capacity $r^2$ as a function of the input standard deviation $\sigma_{\text{in}}$ for the error function activation.
As shown in Fig.~\ref{fig_activation_noise_scaling}(b), the rescaled curves exhibit the same scaling behavior as those obtained with the $\tanh$ activation, characterized by a scaling exponent $\eta = 1/3$.
Figures \ref{fig_activation_noise_scaling}(c) and (d) show $r^2$ as a function of $\sigma_{\text{in}}$ and its rescaled counterpart, respectively, for the fifth-order activation function.
In this case, the observed scaling exponent is $\eta = 1/5$, reflecting the higher-order nonlinearity of the activation function.
Figure \ref{fig_activation_noise_scaling}(e) shows $r^2$ for the piecewise linear activation function, where the capacity remains independent of $\sigma_{\xi}$ when it is sufficiently small, indicating that $\eta = 0$. 
These results support the hypothesis that the scaling exponent $\eta$ is determined by the inverse of the leading-order nonlinearity $\zeta$ in the expansion of the activation function, i.e., $\eta = 1/\zeta$.

\section{Influence of round-off error}
\label{sec:round-off_error}

%%%%%%%%%%%%%%%%%%%%%%%%%%%%%%%%%%%%
\begin{figure}
\centering
\includegraphics[width=8.6cm]{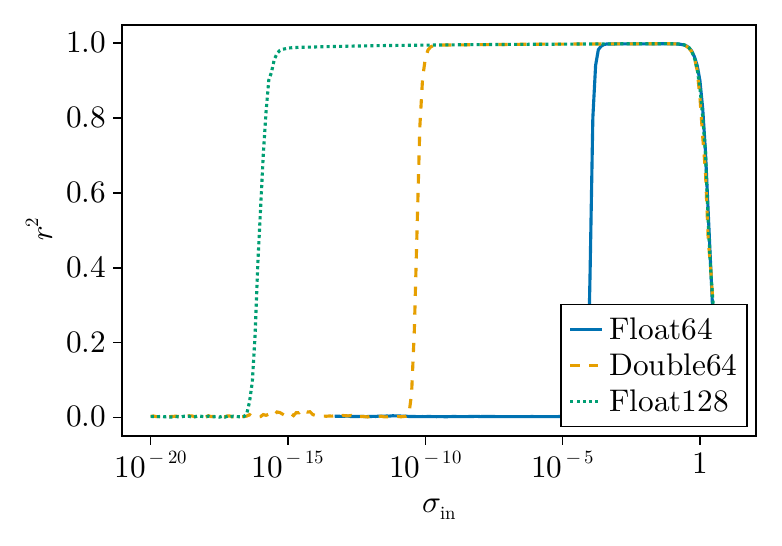}
\caption{Information processing capacity $r^2$ as a function of $\sigma_{\text{in}}$ for various floating-point formats.
The number of nodes is \(N = 200\), and the spectral radius of \(W\) is set to \(\rho = 0.6\).
The noise intensity \(\sigma_{\xi}\) is zero.
The XOR task is performed with a delay of \(k = 7\).
The floating-point formats used are Float64 (significand: 53 bits, exponent: 11 bits), Double64 (significand: 106 bits, exponent: 11 bits), and Float128 (significand: 113 bits, exponent: 15 bits).
As the numerical precision increases, the transition from the linear to the nonlinear information processing regime shifts to smaller values of \(\sigma_{\text{in}}\).}
\label{fig_rounding_error}
\end{figure}
%%%%%%%%%%%%%%%%%%%%%%%%%%%%%%%%%%%%

In this appendix, we discuss the influence of round-off error in numerical calculations.  
Let us consider the case without external noise, i.e., $\sigma_{\xi} = 0$.  
As shown in Eq.~\eqref{transition_noise_scaling}, the critical input standard deviation $\sigma_{\text{in},0}$ vanishes in the limit of large network size $N$.
However, we find that even when $\sigma_{\xi} = 0$, the transition between linear and nonlinear information processing regimes still occurs at a small but finite value of $\sigma_{\text{in}}$.  
We attribute this residual transition to round-off error in the numerical calculations.  
Such numerical error effectively acts as a source of noise, introducing an implicit $\sigma_{\xi}$ that prevents the nonlinear regime from extending to $\sigma_{\text{in}} = 0$.  

To test this hypothesis, we calculate the information processing capacity $r^2$ using different numerical precisions: 
\begin{enumerate}
\item Float64 (significand: 53 bits, exponent: 11 bits).
\item Double64 (significand: 106 bits, exponent: 11 bits).
\item Float128 (significand: 113 bits, exponent: 15 bits).
\end{enumerate}
All calculations, including the network evolution and linear regression, are performed using the respective floating-point formats specified above.
Figure~\ref{fig_rounding_error} shows $r^2$ as a function of the input standard deviation $\sigma_{\text{in}}$ for each precision level.
We observe that, as numerical precision increases, the critical input standard deviation $\sigma_{\text{in},0}$ at which the transition occurs shifts to smaller values.
This result supports the conclusion that the observed transition in the absence of noise ($\sigma_{\xi} = 0$) is an artifact induced by numerical round-off error.

\end{document}